# Surface-Emitting Resonator Interference Microscopy for Label-Free Monitoring of Membrane Dynamics

Chaoyang Gong[1,2], Yu-Cheng Chen[2,*]

[1] Key Laboratory of Optoelectronic Technology and Systems (Ministry of Education of China), School of Optoelectronic Engineering, Chongqing University, Chongqing 400044, China.

[2] School of Electrical and Electronic Engineering, Nanyang Technological University, Singapore 639798, Singapore

Correspondence Email: yucchen@ntu.edu.sg

**Abstract**

Cellular membrane dynamics play an important role in a variety of physiological processes. However, due to the stringent light-coupling conditions required for exciting evanescent waves, label-free mapping of cellular membrane dynamics on curved substrates remains challenging. Here, we report surface-emitting resonator interference microscopy (SERIM), which employs the evanescent wave naturally present in the near-field region of a whispering gallery mode (WGM) resonator to probe the subcellular membrane dynamics. The WGM resonator provides strong optical feedback for enhancing the light-mattering interaction and also provides a biomimetic curvature interface to investigate the membrane dynamics. The interaction of the evanescent wave with the cell membrane caused significant scattering, forming a highly sensitive interference pattern. We found that the time-resolved interference patterns can be utilized to extract subcellular membrane dynamics with a diffractive limited spatial resolution (~ 1 μm). We further employed the SERIM to investigate the spatial heterogeneity of membrane dynamics during migration, and also stimulus responses to temperature and drugs. In striking contrast to the conventional label-free methods, the easy excitation of the evanescent wave makes SERIM a versatile label-free strategy for studying cellular behavior on a curved surface. Our work holds promise for sensitive detection of subtle biochemical and biophysical information during cell-substrate interaction.

# 1. Introduction

Cells sense and respond to their environments by constantly regulating the membrane morphologies[1,2]. As a result, the cell membrane is highly dynamic and plays an important role in a variety of cellular processes such as cell migration, adhesion, differentiation, cell division, and mechanoresponse[3,4]. Membrane dynamics can be used as an inherent indication with a high association with pathological conditions[5]. Robust and nondisruptive methodologies for profiling membrane dynamics at the sub-cellular level and mapping changes after exposure to external stimulus are critical in understanding the cellular physiological activities and the pathogenesis of diseases[5-9].

Imaging the dynamics of cell membranes directly through optical microscopy poses a considerable challenge due to the nanoscale membrane structures falling below the resolution limit of the optical system[10]. As a cornerstone of research in membrane dynamics, fluorescence-based imaging technologies have significantly boosted the tremendous expansion of cell biology[10-15]. However, traditional fluorescence-based technologies suffer from a strong photobleaching effect and fluorescence background due to the overlay of signals out of the focal plane[16,17]. Despite label-free optical imaging techniques, such as total internal reflection microscopy (TIRM)[18], surface plasmon resonance microscopy (SPRM)[19], and reflection interference contrast microscopy (RICM)[20] can reveal cell-substrate interaction with high spatial resolution, issues have also been raised during sensing application to ubiquitous micron-sized curved topography in complex three-dimensional extracellular environments. The TIRM and SPRM employ the evanescent wave to probe the tiny changes in the cell membrane. The evanescent wave excitation has strict requirements on incidence angle, which can hardly be achieved on a curved substrate[21]. The interference pattern generated from RICM strongly depends on the phase difference between the reflective light[22]. The presence of a curved surface induces an additional phase difference, thereby complicating the reliable retrieval of spatial information about membrane dynamics. Over several decades, numerous studies have suggested that substrate curvature influences cell behavior including morphology, polarity, and functions[23,24]. Therefore, developing a label-free imaging technique capable of measuring membrane behavior on a biomimetic curve in a fluorescent label-free manner is of paramount importance for better understanding the interaction between cells and the environment.

Here, we report surface-emitting resonator interference microscopy (SERIM) for directly mapping the membrane dynamics on a whispering gallery mode (WGM) resonator. As illustrated in Figure 1a, the blood vessel-like cylinder structure of optical fiber provides a curved surface that mimics the in vivo microenvironment for investigating cellular behavior. Meanwhile, the optical fiber serves as an extremely high-Q resonator supporting WGM, which significantly amplifies the subtle changes in the cavity interface. Once filled with organic dye and pumped with a pulsed nanosecond laser, the optical fiber generates laser emission, thus enabling a strong evanescent wave interaction with the cell membrane. The scattered evanescent waves interfere with each other and form a unique interference pattern (Fig. 1b). As evanescent waves decay exponentially with distance from the fiber surface, the interference pattern can reveal cell membrane dynamics in the near-field region without influence from cellular activities deep in the cytoplasm (Fig. 1c). Our finding indicates that the profile of the interference pattern agrees well with the cell morphology and the membrane activity can be mapped with high sensitivity and high resolution (Figs. 1d to 1f). Our work provides a versatile label-free solution for quantitatively characterizing cell membrane dynamics on a curved surface.

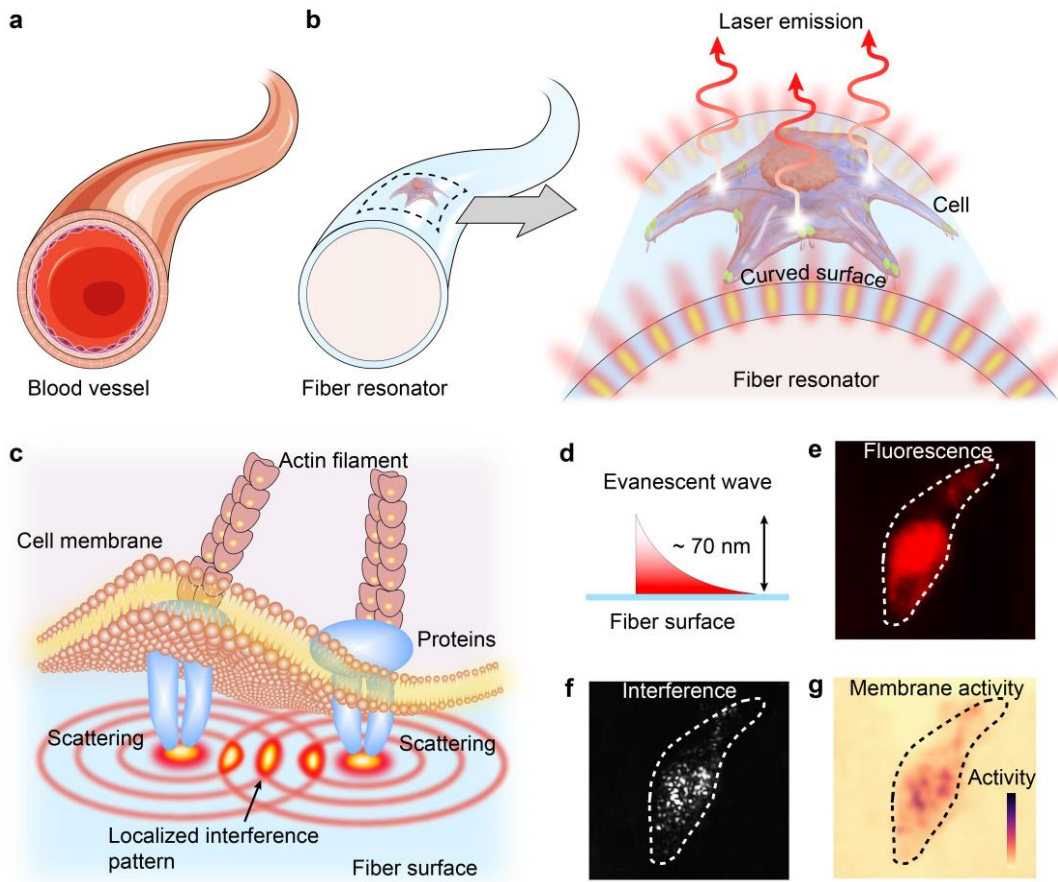

**Fig. 1 Concept of surface-emitting resonator interference microscopy (SERIM).** (**a**) Blood vessel. (**b**) Conceptual illustration of imaging membrane dynamics on a lasing WGM resonator. Left panel: Illustration of fiber resonator, which has a similar cylinder structure to a blood vessel. Right panel: Enlargement of the area in the boxed region on the left. (**c**) Illustration of the interaction between evanescent wave and cell membrane. The interference of scattered light forms localized interference patterns. (**d**) Illustration of the evanescent wave on the fiber surface. (**e,f**) Fluorescence image (**e**) and interference pattern (**f**) of a cell. (**g**) Reconstructed spatial distribution of the membrane dynamics.

## 2. Results
### 2.1 Principle of SERIM

As a conceptual demonstration, we investigated the light-matter interaction on the fiber surface with gold nanoparticles (AuNPs) before applying biological cells to the optical fiber. Figure 2a shows the image of optical fiber. The customized optical fiber was fabricated with the fiber draw tower, and it was used as a microresonator to generate laser emission (Supplementary Figs. 1 to 2). As shown in Fig. 2a, the optical fiber has a hollow cylinder geometry with a diameter of 136 μm and a thickness of 4 μm, which can be easily adjusted in the fabrication process. The fiber drawing process involves a melting procedure that significantly enhances surface quality, enabling the optical fiber to support a high Q-factor ($> 10^6$) WGM[25]. The strong evanescent wave of the WGM leaks onto the outer surface of the optical fiber, allowing for a strong interaction light-matter interaction. Our simulation shows that a single AuNP with a diameter of 20 nm binding on the fiber surface will couple a fraction of light out of the resonator through scattering, thus enabling the detection of the single nanoparticle binding events through far-field imaging (Fig. 2b).

Low threshold laser emission was observed from the dye-filled fiber resonator. When the pump energy density is below the lasing threshold, only weak fluorescence emission located at the pump location is observed (Dashed circles in Supplementary Fig. 3). Once the pump energy density exceeds the lasing threshold, a bright laser rim appears on the fiber boundary. Meanwhile, a series of sharp peaks corresponding to the longitudinal modes can be observed in the emission spectrum, which confirms the successful generation of laser emission (Fig. 2c). Thanks to the strong light-matter interaction induced by the high Q-factor, the dye-filled optical fiber exhibit a low lasing threshold down to 1.4 μJ/mm$^2$ (Fig. 2d). This low energy density is safe for most biomaterials including molecules, cells, tissues, and living organisms[26-31]. Different from the unidirectional fluorescence emission, the WGM laser emission escapes tangentially from the fiber boundary, which contributes to a very high contrast image. As illustrated in the inset of Fig. 2d, once a single AuNP is bound on the fiber surface, a bright spot with a diffraction-limited size (~ 1.1 μm in diameter, See Supplementary Fig. 4 for more details) can be observed due to scattering. Note that the bright spot is only observable when the pump energy density exceeds the lasing threshold. This phenomenon suggests a significant increase in scattering due to enhanced light-matter interaction on the lasing fiber surface.

The SERIM can also provide time-resolved images that reveal nanoparticle binding dynamics. Figure 2e shows the dynamic evolution of scattering imaging when a single AuNP is binding onto the fiber surface. The gradually increasing scattering intensity indicates a closer distance to the fiber surface. We extracted the scattering intensity and the XY position of the AuNP in Figs. 2h and 2i, respectively. At t = 0 s, the AuNP lies out of the evanescent wave and no scattering light can be observed. Once the AuNP enters the evanescent wave region, the interaction of AuNP with the optical field results in scattering, and a small spot with weak intensity appears in the image (t = 17 s). With the distance to the optical fiber getting closer, the AuNP experiences an exponentially stronger evanescent field, and a brighter scattering image was observed (t = 23.1 s). Due to Brownian motion, the AuNP shows a blinking intensity (Fig. 2h) and a stochastic motion (Fig. 2i). When the AuNP is finally attached to the fiber surface, the stochastic motion stops and a maximum scattering intensity is observed (t = 34 s). The dynamic light scattering signal obtained by SERIM can potentially used for quantificating the size and concentration of molecules in solution[32].

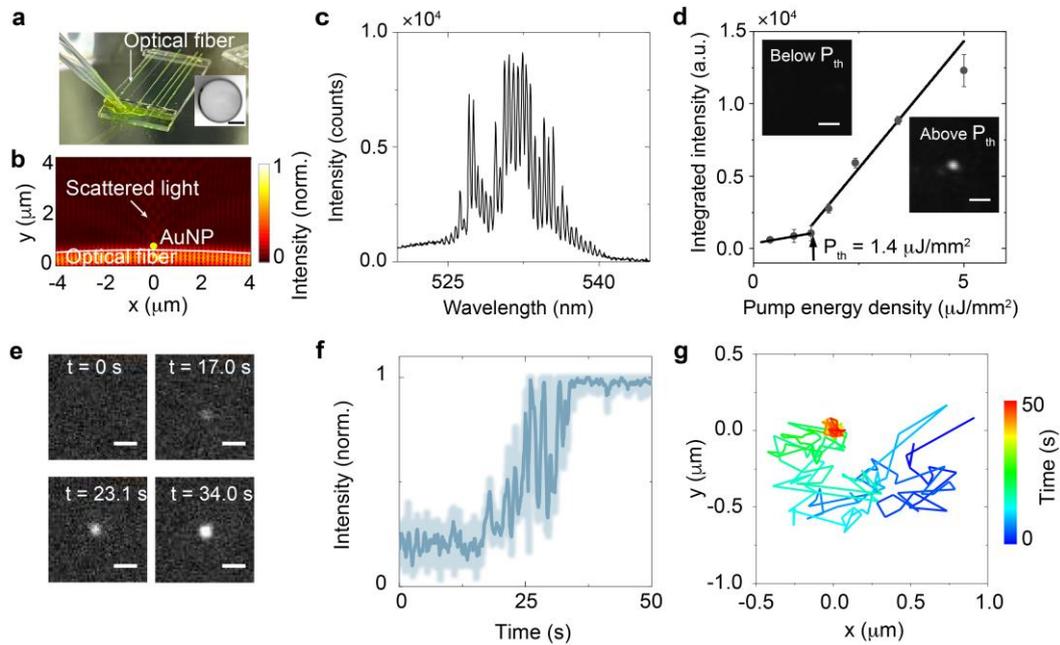

**Fig. 2 Validation of SERIM with a single Au nanoparticle (AuNP).** (**a**) Photo of dye-filled optical fiber. Inset, a microscopic image of hollow fiber cross-section. Scale bar: 50 μm. (**b**) Simulation of light scattering from fiber surface induced by 20 nm AuNP. (**c**) Emission spectra above lasing threshold. (**d**) Threshold curve of the dye-filled optical fiber. Inset, SERIM image of a single AuNP with pump energy density below

and above the lasing threshold. Scale bar: 3 μm. (**e,f,g**) Time-resolved scattering images (**e**), scattering intensity (**f**), and spatial location (**g**) of a single AuNP during binding on the fiber surface. Scale bar: 3 μm.

**2.2 Label-free imaging of membrane dynamics**

We next demonstrated that the SERIM can label-free imaging reveal subtle changes in the membrane. A muscle tissue cell line C2C12 was seeded on the optical fiber overnight to enable stable attachment on the resonator surface (See Methods for details). As shown in Fig. 3a, the optical fiber shows good cell compatibility. The cells proliferate from a density of $7.9 \times 10^3$ cells/cm$^2$ to $1.6 \times 10^4$ cells/cm$^2$ after two days of incubation. Different from the scattering image of a single AnNP in Fig. 2e, the SERIM image of a cell shows a unique interference pattern with high contrast (See Supplementary Fig. 5 for more details). As illustrated in Fig. 1c, a fraction of the intracavity field was coupled out by the cell membranes through scattering and interference with each other. The resulting interference pattern (dashed curves) shares a similar profile with the bright-field (BF) image, yet it also reveals more subtle substructures with periodic bright spots (Fig. 3b). Because of the strong light-matter interaction on the fiber surface, the tiny changes in the cell membrane will induce a phase change of the scattered light, resulting in a significant change in the interference pattern. A dynamic interference pattern was observed in a live cell (Fig. 3c), while the pattern remained stable for a fixed cell (Supplementary Fig. 6). According to our calculation in Supplementary Eq. 1, the evanescence wave is highly localized with a penetration depth of about 73 nm[33], which matches the typical range of cell-substrate distances (~ 40 to 60 nm)[19]. Therefore, the interference pattern can only respond to the behavior of the cell membrane in the near-field region. Once the cell membrane dynamics were inhibited, a stable interference pattern similar to that of the fixed cell was observed (Supplementary Fig. 7). This result confirms that the changes in the interference pattern are sensitive to the membrane dynamics while being insensitive to the intracellular activities in the cytoplasm.

We also attempted to extract the spatial heterogeneity of membrane dynamics. The spatial distribution of membrane activities from the time-resolved interference patterns (See Methods for more details). Briefly, we recorded the intensity evolution of each pixel, and the results of 3 representative pixels (A, B, and C in Fig. 3c) are shown in Fig. 3d. The temporal evolution of each pixel shows significantly different trends, which is caused by the spatial heterogeneity of membrane dynamics. Then, the membrane activity of each pixel can be quantified by using Eqs. 1 to 2, and the spatial distribution of the membrane activity can be mapped. An active region (in yellow color) corresponding to the highly active areas (Fig. 3e) can be identified with a spatial resolution that approaches the diffraction limit (Fig. 3f). The spatial resolution can be further improved by employing an objective with a larger numerical aperture (NA). The ability to map subcellular membrane dynamics with high spatial resolution makes SERIM a powerful tool for cell biology.

The SERIM has significant advantages over the conventional evanescent wave-based method in revealing membrane dynamics. Firstly, light coupling in a fiber resonator is much easier than the conventional methods. During the lasing process, the pump laser excites the organic dye molecules from the far field, causing the dye molecules to emit photons in all directions. A portion of these photons that meet the resonance conditions automatically couples into the whispering gallery mode (WGM) and is subsequently amplified by stimulated emission. As a comparison, efficient excitation of evanescent waves in the conventional TIRM and SPRM has strict requirements on incidence angle. Secondly, the hollow optical fiber provides a curved surface that mimics the in vivo microenvironment for investigating cellular behavior. Limited by light coupling conditions, the conventional evanescent wave-based methods can hardly be applied on curved surfaces. Thirdly, pulsed laser excitation in SERIM greatly reduces the thermal effects of continuous wave widely used in conventional methods, which could otherwise affect the dynamic measurement. In our experiment, an average excitation power as low as 75 nW ( 1.5 nJ @ 50 Hz) is sufficient for clear image acquisition, which is 4 orders of magnitude lower than the typical value (~ 1 mW) used in TIRM and SPRM.

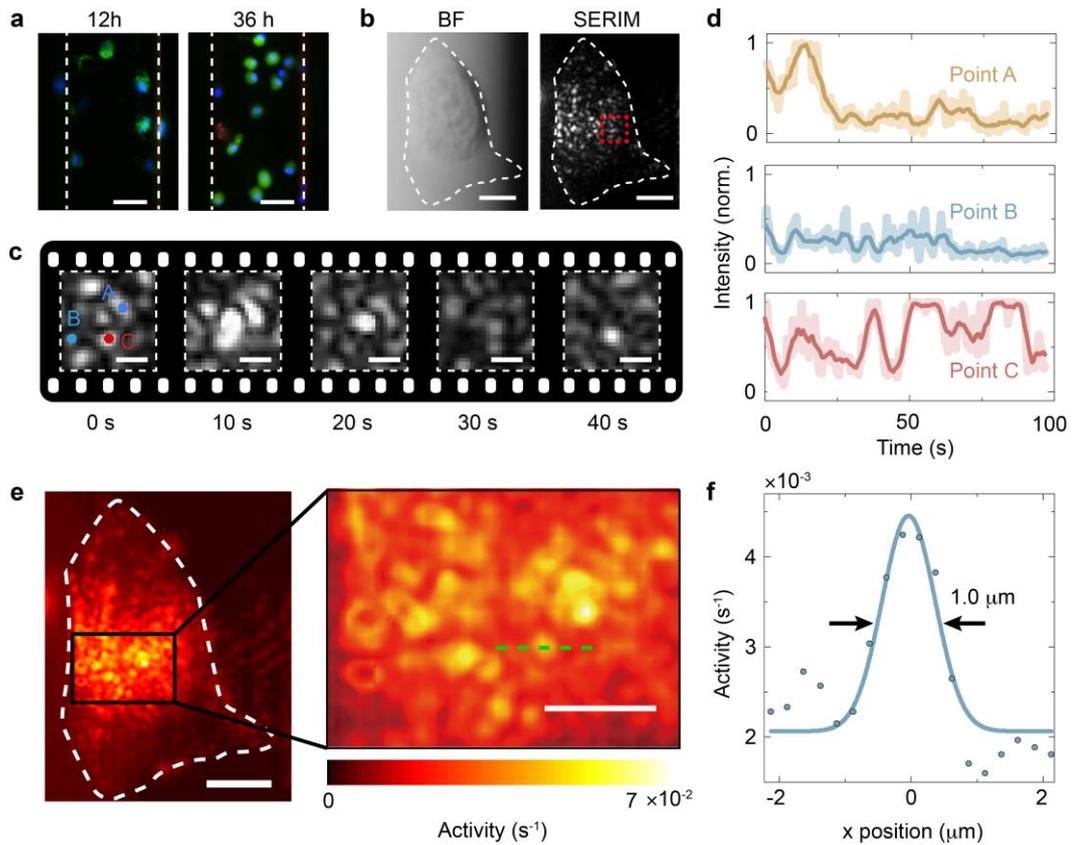

**Fig. 3 Label-free imaging of membrane dynamics.** (**a**) Live/dead staining of cells on optical fiber. Red, dead cells; Green, live cells; Blue, cell nucleus. The dashed lines indicate the edges of the optical fiber. Scale bar: 50 μm. (**b**) Comparison of the bright field (left) and the SERIM image (right) of a cell. Scale bar: 10 μm. (**c**) Time-resolved SERIM images. These images are cropped from the boxed area in the right panel of (**b**). Scale bar: 2 μm. (**d**) Temporal evolution of the intensity of points A, B, and C. (**e**) Spatial distribution of membrane activity. Scale bar: 10 μm. Inset, enlargement of images in the boxed region. Scale bar: 5 μm. (**f**) The spatial resolution of SERIM. Data are extracted along the dashed line in the inset of (**e**).

## 2.3 Subcellular membrane dynamics during migration

Cell migration is a fundamental biological process associated with membrane dynamics including leading-edge formation and trailing-edge retraction[34,35]. It plays a central role in a wide variety of biological phenomena, such as embryonic morphogenesis, immune surveillance, and tissue repair and regeneration[36]. Aberrant regulation of cell migration drives the progression of many diseases, including cancer invasion and metastasis[37,38].

Hence, studying the dynamics of cell membranes during migration is crucial for understanding pathological development.

To demonstrate the real-time capability of SERIM, we recorded the interference images during cell migration and reconstructed the live maps of membrane activity (See Methods for more details). Figures 4a and 4b show the time-elapse images and the reconstructed activity map of a lung carcinoma epithelial cell (A549). The spatial distribution of the activity map is highly dynamic, containing rich information on the cellular behavior that the conventional method cannot reach. To better describe the temporal evolution of the membrane activity map, we define a center-edge (CE) vector, which starts from the center of the cell and points toward the farthest edge of the active region (indicated as arrows in Fig. 4b). As illustrated in Fig. 4c, because of the guidance effect of a cylindrical surface[23], the cell shows a dominant migration along the fiber axis. Interestingly, the CE vector agrees well with the migration direction. This phenomenon is correlated with the coordination of membrane dynamics during cell migration, which involves delivering a new membrane to the leading edge and internalizing adhesive receptors from the cell rear[39]. We plotted the activity profile along the direction of the CE vector in Fig. 4d. At t = 0 min, the activity profile is symmetric about the cell center. The profile started to spit into two symmetric peaks at t = 10 min. Then, the left peak disappeared and remained only the peak in the leading edge of the migration direction. We believe the results could uncover how cells perceive the environment and regulate their membrane dynamics during migration. For example, when cells are "deciding" the migration direction, they may sense their environment in different directions (the split two peaks observed in Fig. 4d.). Once they have "chosen" their direction, the membrane activity on the opposite side decreases while the membrane in the leading edge remains active to help migration and sensing the environment.

The heterogeneity of membrane dynamics in different cell lines is also evaluated. We tested four types of cell lines, including C2C12, A549, Panc-1 (pancreatic cancer), and MCF-7 (breast cancer) cell lines (Fig. 4e). Because different cell lines show significantly different behavior and functions, their membrane activity is heterogeneous.

Panc-1, for example, is highly invasive and metastatic with the highest membrane activity. MCF-7, on the other hand, is non-invasive and has the lowest activity. This result indicates that our method can easily distinguish the differences in membrane activity in different cell lines.

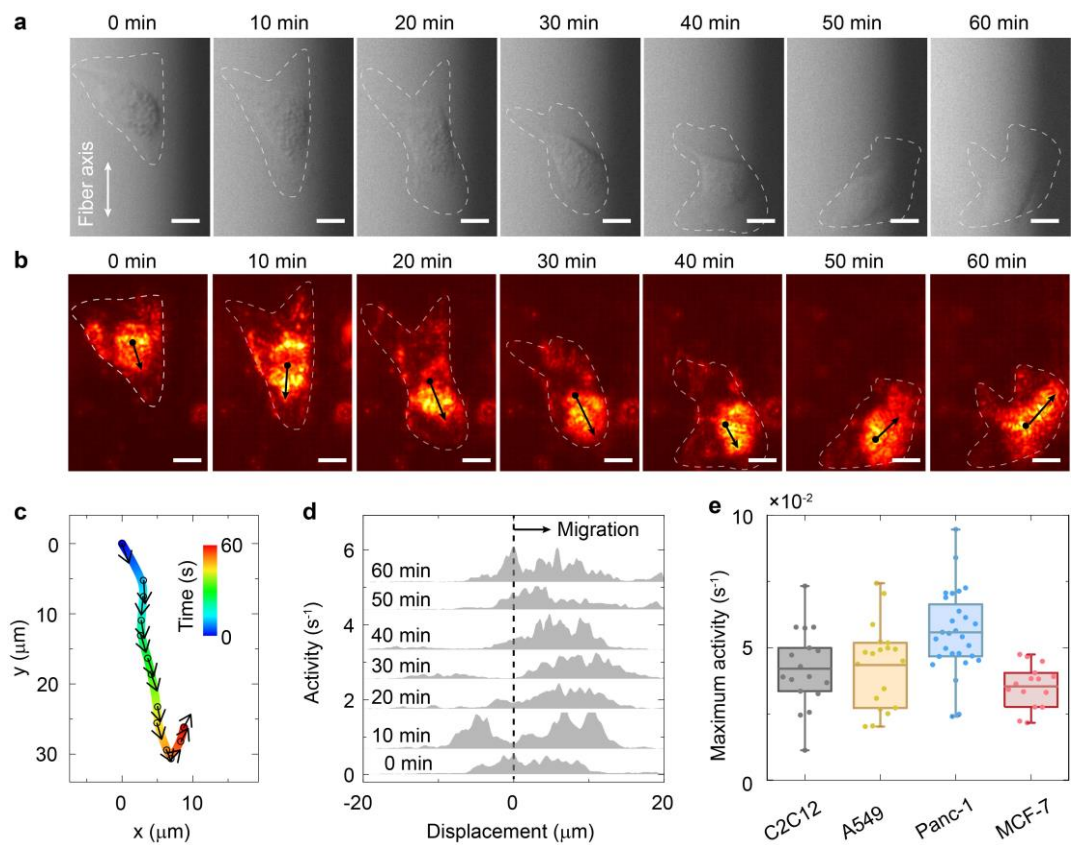

**Fig. 4 Monitoring membrane dynamics during migration.** (**a, b**) Time-resolved bright-field images (**a**) and membrane activity maps (**b**). Scale bar: 10 μm. The black arrows represent the CE vectors. (**c**) Tracking of cell migration on the curved resonator surface. Black arrow, the CE vectors. (**d**) Profile of membrane activity along CE vectors at different observation times. (**e**) Tukey boxplot of membrane activity of different cell lines.

### 2.4 Stimulus-response of membrane dynamics

Cell membranes serve as an interface between the cell and its environment, which can sense and respond to the stimulus from the surrounding environment. We further employed the SERIM to investigate the stimulus-response of membrane dynamics to temperature and drugs.

Temperature influences membrane activities by changing the membrane fluidity. At lower temperatures, the membrane fluidity decreases because the fatty acid tails of the phospholipids become more rigid[40]. Hence, a lower membrane cavity can be expected at a lower temperature. We compared the activity map when the temperature was reduced from 37ºC to 0ºC and then returned to 37ºC. As illustrated in Fig. 5a, the activity map shows a strong temperature dependence. The maximum activity decreased to 62% of its original value when the temperature was lowered from 37 °C to 0 °C, and subsequently recovered to 90% of its original value when the temperature was raised to 37°C (Fig. 5b).

We also investigated the stimulus-response of cells to drugs by monitoring the subcellular membrane dynamics. As illustrated in Fig. 5c, Cytochalasin D (Cyto-D) is a specific inhibitor of actin polymerization, which plays a significant role in cell motility. β-lapachone (β-Lap) is an anti-cancer drug that induces cytotoxicity by modulating reactive oxygen species. Figures 5d and 5e show the activity map after Cyto-D and β-Lap treatment, respectively. A gradual decrease in membrane activity was observed, while no significant migration or changes in the spatial distribution of the activity map were detected. The influence of the two drugs on membrane activity is shown in Fig. 5f, indicating a significant inhibition of membrane activity. Similar results can also be observed after treatment with Blebbistatin (Blebb.) and Y27632. Here, Blebb. is an inhibitor of myosin molecular motor that converts chemical energy in the form of ATP to mechanical energy. Y27632 is a selective inhibitor of the Rho-associated protein kinase, which involves mainly regulating the shape and movement of cells. Compared with the control group (without drug), all these drugs reduced membrane activity significantly (Fig. 5g). The successful investigation of the cellular response to various drugs suggests that SERIM also holds potential for application in drug research.

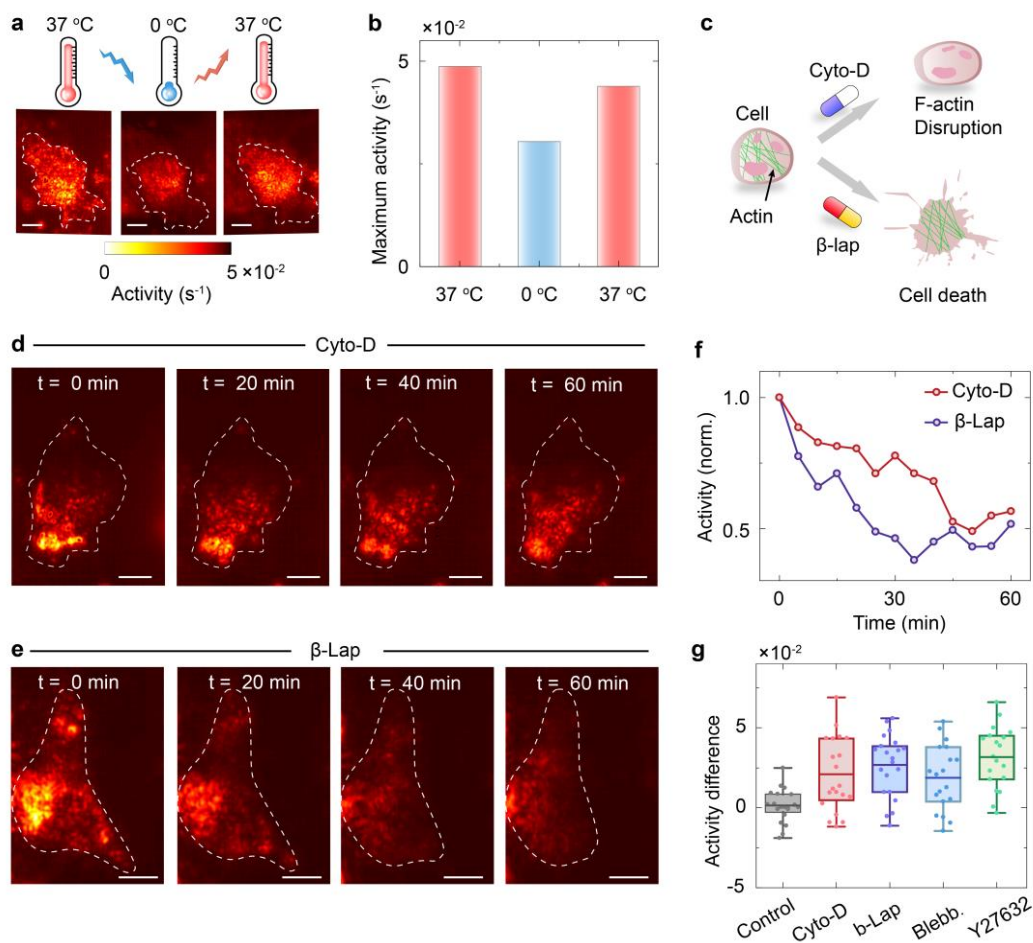

**Fig. 5 Stimulus-response of cells.** (**a, b**) The activity map (**a**) and the maximum activity (**b**) under different temperatures. Scale bar: 10 μm. (**c**) Effect of Cyto-D and β-Lap on cell membrane dynamics. (**d,e**) Time-resolved activity map with Cyto-D (**d**) and β-lap (**e**) treatment. Scale bar: 10 μm. (**f**) Temporal evolution of the activity after drug treatment. (**g**) Tukey boxplot of activity difference.

## 3. Discussion

Optical imaging technologies have made tremendous milestones in gaining physical and chemical understandings of biological systems. However, most of the present label-free imaging technologies are designed for flat surfaces, which cannot represent the curved surfaces encountered in vivo. Conventional label-free imaging technologies such as TIRM, SPRM, and RICM face challenges in providing accurate information on micro-scale curved surfaces due to strict coupling conditions and curvature-induced phase differences. The SERIM addresses this issue by using the evanescent wave naturally present in the near-field region of a WGM resonator to detect subtle changes in the cell membrane. The exponentially decaying characteristic of evanescent wave ensures a strong light-matter interaction with the cell membrane, rather than with intracellular activities in the cytoplasm. The strong interaction between the cell membrane and the evanescent wave leads to a significant scattering of light from the microresonator, generating a highly sensitive interference pattern with strong contrast. We have demonstrated that the temporal evolution of interference patterns can be utilized to extract valuable insights into membrane dynamics and the stimulus-response to the environment. This concept employs optical resonators to mimic the curvature surface in vivo environment, providing a versatile platform for studying cell behaviors.

We envision that the SERIM could potentially be applied for various biochemical and biophysical applications. Firstly, SERIM could enable operando monitoring of cellular secretions at the single-molecule level, particularly for protein molecules. Recent studies have shown that an individual protein molecule can produce a detectable scattering signal[41]. Specific detection can be achieved by functionalizing the fiber surface with antibodies. This approach could also be adapted for the detection of single vesicles. Secondly, subtle information on cell-surface interactions could also be investigated by coating a thin layer of extracellular matrix (ECM) to the fiber surface. The interaction between the cell and ECM will alter changes in the microscopic structure of the fibril network, thus changing the interference pattern. Mechanical information on weak cell-substrate interactions, such as contraction forces, could potentially be detected from the temporal evolution of these interference patterns.

**4. Methods**

*4.1 Optical system setup*

An upright microscope system with ×10 objectives (NA = 0.3) is used for excitation and signal collection (Fig. S1). A nanosecond pulsed laser (EKSPLA NT230, 50 Hz, 5 ns pulse width) integrated with an optical parametric oscillator was used as an optical pump. The pump laser (450 nm) was focused on the optical fiber resonator with a diameter of 32 μm. The laser emission was sent to a spectrometer (Andor Kymera 328i) and a scientific complementary metal-oxide-semiconductor (sCMOS) camera, respectively. The spectrometer and the sCMOS records the spectra and laser image with a frame rate of 2.5 frame/s, respectively. A bandpass filter was added in front of the beam splitter (BS) to eliminate the influence of the pump laser on the signal collection. For the cell experiment, a stage top chamber (okolab, H301) was used to keep the temperature at 37 ºC.

*4.2 Preparing of the fiber resonator*

The optical fiber (~ 3 cm in length ) was immersed in acetone to remove the polymer coating. The fiber resonators were integrated as an array on a customized glass substrate (Supplementary Fig. 1) with UV glue. After that, the optical fiber was treated with UV-ozone cleaner (Ossila, L2002A2) for 1 h to increase the hydrophilicity. 20 mM C6 dye solution was prepared by dissolving the C6 powder (Sigma, No. 442631) in immersion oil ($n$ = 1.52) and subsequently was withdrawn into the optical fiber to serve as a gain medium.

*4.3 Monitoring the binding dynamics of AuNP*

The dye-filled fiber resonator was immersed in PBS. AuNP (Sigma, No. 741965) was diluted 1000 times and added to the solution. The binding dynamics were monitored by recording the time-resolved scattering images.

*4.4 Cell culture*

All the cells including C2C12, A549, Panc-1, and MCF-7 were cultured in Dulbecco's Modified Eagle Medium (DMEM) (Gibco, No. 11054020) supplemented with 10% fetal bovine serum (FBS) (Gibco, No. 10099158) and 1% penicillin-streptomycin (Gibco, No. 10378016) mixed solution. This formulation contains no phenol red which could probably affect the laser measurement.

The dye-filled optical fiber was first sterilized in 70% (v/v) ethanol solution. After 5 times washing with phosphate-buffered saline (PBS). Then, the sterilized fiber resonator was transferred into a petri dish and immersed into the cell culture medium. After that, the cells were seeded on the optical fiber and cultured for 24 hours in a humidified atmosphere of 95% air and 5% $CO_2$ at 37 ºC and were ready for further experiment.

*4.5 Reconstructing the spatial distribution of membrane activity*

We recorded the temporal evolution of the interference patterns frame by frame. Each frame can be written as a matrix

$$A(t) = \begin{pmatrix} a_{11}(t) & \cdots & a_{1n}(t) \\ \vdots & \ddots & \vdots \\ a_{m1}(t) & \cdots & a_{mn}(t) \end{pmatrix}, \tag{1}$$

with $a_{ij}(t)$ denoting the intensity temporal evolution of a single pixel at the location of $(i, j)$. For each pixel, we firstly normalized the temporal signal by using $A_{ij}(t) = a_{ij}(t)/A_{max}$. $A_{max} = \max(A(t))$ is the maximum intensity recorded during the whole observation time. Then, we calculated the moving average by using $\overline{A_{ij}}(t) = \sum_{k=0}^{w-1} A_{ij}(t+k)/w$, where $w = 10$ is the window length. The activity of each pixel is defined as

$$\gamma_{ij} = \frac{1}{T}\int_0^T |\overline{A_{ij}}(t)'| dt. \tag{2}$$

Here, $\overline{A_{ij}}(t)'$ denotes the first-order derivative of the moving average. $T$ is the length of observation time. Mapping the activity at the subcellular level was achieved by employing Eq. 2 for each pixel in the interference pattern.

*4.6 Real-time mapping membrane dynamics during migration*

The activity map was obtained with an interval of 5 min. For each measurement, the dynamic interference patterns were collected for 60s with a frame rate of 2.5 frame/s. The obtained interference patterns were further converted activity map with Eqs. 1 to 2.

*4.7 Recording the stimulus-response of cell membrane*

To record the stimulus response to temperature, we first recorded the membrane dynamics at 37°C in a stage top chamber. To induce temperature stimulus, the chamber was switched off and the petri dish was put into the ice/water mixture (0 °C) for 40 minutes. After recording the membrane dynamics at 0 °C, we turned on the chambers and the temperature was recovered from 0 °C to 37 °C. The fiber resonator was incubated at 37 °C for another 40 min before laser recording.

To record the stimulus-response to drugs, we induced Cyto-D (1 μM), blebbistatin (50 μM), Y27632 (10 μM), and β-lap (1 μM) in the cell culture medium and incubated for 1 hour at 37°C before laser recording. Then, we recorded the temporal evolution of the subcellular dynamics with an interval of 5 minutes.

**Acknowledgments**

The authors thank Prof. Tong Ling (Nanyang Technology University, Singapore) for the discussion. This research is supported by A*STAR MTC IRG-Grant (M21K2c0106, Singapore).

**Author Contribution**

Y.-C.C. formulated the idea behind the study. C.G. and Y.-C.C. designed experiments. C.G. performed the experiments and wrote the manuscript. C.G., G. F., S. Z., and Z.Q. performed data analysis. C.G. and J.X. developed the analysis software. G.Y provided cells. G.-D.P. fabricated optical fiber. C.G., Y.G., and Y.-C.C revised the manuscript.

**Conflict of interest**

All the authors declare no conflict of interest.